\def\year{2019}\relax
\documentclass[letterpaper]{article} 
\usepackage{aaai19}  
\usepackage{times}  
\usepackage{helvet}  
\usepackage{courier}  
\usepackage{url}  
\usepackage{graphicx}  
\usepackage{todonotes}  
\begin{document}
%
\title{Taking the Scenic Route: Automatic Exploration for Videogames}
\author{Zeping Zhan, Batu Aytemiz, Adam M. Smith\\
Design Reasoning Lab \\
University of California, Santa Cruz\\
{\{zzha50, baytemiz,amsmith\}@ucsc.edu}
}
\maketitle
\begin{abstract}

Machine playtesting tools and game moment search engines require exposure to the diversity of a game's state space if they are to report on or index the most interesting moments of possible play. Meanwhile, mobile app distribution services would like to quickly determine if a freshly-uploaded game is fit to be published. Having access to a semantic map of reachable states in the game would enable efficient inference in these applications. However, human gameplay data is expensive to acquire relative to the coverage of a game that it provides. We show that off-the-shelf automatic exploration strategies can explore with an effectiveness comparable to human gameplay on the same timescale. We contribute generic methods for quantifying exploration quality as a function of time and demonstrate our metric on several elementary techniques and human players on a collection of commercial games sampled from multiple game platforms (from Atari 2600 to Nintendo 64). Emphasizing the diversity of states reached and the semantic map extracted, this work makes productive contrast with the focus on finding a behavior policy or optimizing game score used in most automatic game playing research.

\end{abstract}

\noindent To know what's inside of a videogame, you need to do more than read source code or browse asset data. Looking over a player's shoulder can give you one view of a game's space of possible interactions, and running an artificial intelligence (AI) algorithm to find score-optimizing behavior can give you another. Neither, however, is trying to play in a way that intentionally covers the most ground.

Access to a large and diverse sample of trajectories through a game's state space could enable many applications relevant to game designers, players, scholars, educators, and distributors. We might automatically map reachable zones of a game's world \cite{mappy,treefrog} and visualize how these change in response to design modifications. We might index the moments demonstrated and return them in response to visual queries \cite{crawling}, enabling users of a search engine to bookmark and share references to significant events \cite{gisst}. We might even identify when a game freshly uploaded to an app store should be flagged for removal because it behaves maliciously only after several minutes of interaction \cite{malware}.

Meanwhile, in automated gameplay via reinforcement learning (RL), the ability to explore is critical for efficient learning. Aytar et al.~\shortcite{youtube} demonstrate how even a single externally-provided trace of high-reward behavior can be generalized into a reusable behavior policy. We believe that algorithms explicitly designed to address the exploration problem (finding sequences of actions that reach interesting moments by any means necessary) could complement advances in reinforcement learning. 

We introduce the problem of automatically exploring a game's state space with the goal of producing a semantic map (useful for downstream inference tasks) on timescales comparable to human playtesting efforts. This map should allow us to judge the similarity of moments and allow inferences about how moments relate to one another. Automatic exploration is related to the topic of intrinsically-motivated play in reinforcement learning \cite{montezuma}. However, we are interested in the large collection of moments found via exploration rather than training a behavior policy. In particular, we are most interested in uncovering automatic exploration techniques that are plausibly applicable to analysis of contemporary mobile games (native-compiled software binaries for 64-bit computing platforms which draw user interfaces with low-level code) on short timescales (minutes rather than days of in-game experience).

In this paper, we make the following contributions:
\begin{itemize}
\item We define several exploration quality metrics, some of which make use of game-specific knowledge and others that are comparable across games and platforms.
\item We offer a methodology for comparing human and machine exploration efficiency as a function of time.
\item We identify several elementary automatic exploration strategies and demonstrate them on a range of platforms.
\item Our experimental results highlight basic gameplay competency for elementary methods, the ability of our metrics to identify distinct scenes, the ability to bootstrap perceptual models with self-play, and the strength variation for exploration techniques across games and across game platforms spanning from Atari 2600 to Nintendo 64.
\end{itemize}

\section{Background}

This section reviews existing techniques for gameplay knowledge extraction, automated testing of mobile apps, and existing techniques in AI for extrinsically and intrinsically motivated automatic play. 

\subsection{Gameplay Knowledge Extraction}

Our project is situated with the larger effort of automated game design learning (AGDL) \cite{agdl} a research paradigm that encourages understanding games by directly interacting with them (rather than, e.g., source code analysis). In particular, our work is a direct response to a previous paper in the Knowledge Extraction from Games workshop on representing and retrieving game states with \emph{moment vectors} \cite{moments}. The automatic exploration strategies we offer in this paper are a source of gameplay trajectories (screenshots, actions, and memory snapshots) that complements the use of human gameplay data in that work. 

Moment vectors for interactive media are a sub-symbolic knowledge representation similar to that of word vectors in natural language processing \cite{wordanalogy}. Although word vectors can be of direct use in text retrieval applications, they can easily support more complex tasks that rely on the ability to make inferences from a text's semantic content \cite{bagoftricks}. 

Zhang et al.~\shortcite{crawling} describe a visual search engine for game moments that is based on moment vectors derived from various gameplay sources. These authors also note a curious property of moment vectors that we think would make them useful for inference in tasks far beyond retrieval: they support \emph{reasoning by analogy}. Similar to linear algebra analogies in word vector representation, moment vectors learned with the Pix2Mem strategy appear to represent game-specific knowledge such as which power-ups the player has collected or what their location is within the larger game world. 

AGDL systems can only learn from the parts of a game they actually experience. Missing from previous work in this domain is any way of quantifying coverage of a game's content. Although it is difficult to define what it means to see \emph{all} or even \emph{enough} of a game's content, we can still make progress by judging when one exploration method has seen \emph{more} than another. For the first time, the exploration quality metrics contributed in this paper will quantify this level of exposure.

\subsection{App Testing for Mobile Markets}

Mobile app developers continually submit new software for distribution on services like Apple's App Store or Google Play, yielding many thousands of new apps per day. To balance the need for quality control with the financial incentive to bring new apps to market in a timely manner, distributors are pressured to judge whether an app should be published very quickly. In response, scalable techniques have been devised to detect malware or trivial wrappings or repackagings of existing apps using just a few seconds of simulated interaction \cite{seconds}. While this is useful for basic quality control, we believe more interaction is required to gather enough data to support content-based retrieval~\cite{crawling} for apps and the moments they contain.

Either by revenue or by popularity, games stand out as a significant category of mobile apps. Some automated app testing techniques can guess at meaningful actions to try based on parsing graphical user interfaces assembled from platform-provided building blocks \cite{ripping}. However, these techniques break down for many games that present custom interfaces drawn with low-level libraries such as OpenGL ES. Similarly, techniques based on static analysis of bytecode \cite{apposcopy} break down when native machine code is used (e.g.\ in optimized rendering or game physics code). By focusing our exploration of videogames at the level of their display pixels and low-level input events, we adopt a perspective that works across many more game platforms.

\subsection{Extrinsically Motivated Automatic Play}

In most classic applications of AI in games (famously in Chess and Go), the goal is to win the match or attain the optimal score. Monte Carlo Tree Search (MCTS) is one surprisingly simple and effective strategy for finding a sequence of moves in a modeled environment that approximately optimizes this score \cite{mcts} and has already been applied to strategy analysis in games  \cite{zook}. These approaches all require a \emph{forward model} (or simulator) that can list available actions, identify terminal states (with their scores), and (un-)apply actions in support of search. While constructing a fairly accurate forward model for the Atari 2600 using an existing emulator implementation like Stella\footnote{\url{https://stella-emu.github.io/}} is straightforward, the larger memory sizes of recent platforms (e.g.\ Android) complicate snapshot-and-restore.

Model-free reinforcement learning techniques are capable of learning to play videogames without access to this kind of simulator. Playing without the ability to take back an action, Deep Q-Networks were recently shown \cite{dqn} to learn super-human play styles for some Atari games (after more than 38 days of simulated gameplay experience). These results are impressive, but we seek broad state coverage for much more recent games in much less time. Dropping the requirement of learning a search-free action policy allows us to operate under much shorter timescales (minutes rather than days). For our applications, search is an entirely acceptable strategy for reaching new states.

\subsection{Intrinsically Motivated Automatic Play}

\emph{Montezuma's Revenge} is an Atari game that stands out as particularly difficult for many deep reinforcement learning algorithms because the rewards that motivate play (derived from game score increases) occur so sparsely. In response, researchers are starting to add extra rewards representing intrinsic motivations (such as curiosity about infrequently visited states) to existing approaches \cite{montezuma}. Although this might seem to distract from optimizing score, it turns out to help agents discover new rewarding paths. Further work has shown that exclusively using curiosity also leads to favorable results \cite{largescale}. In each of these projects, however, the primary metric used to argue for the effectiveness of a method is to report the score associated with the existing reward metric.

Intrinsic motivation for exploration is not a new topic, even within applications to games \cite{curious}. Other work in AI is specifically oriented towards mapping reachable spaces rather than learning ideal behavior policies. Rapidly Exploring Random Trees (RRT) \cite{rrt} and Probabilistic Roadmaps (PRM) \cite{prm} are two such techniques originally invented for assisting in motion planning for robotics. Bauer et al.~\shortcite{treefrog} introduced RRT to the technical games research community in an application to level design feedback.

Iterative widening is another such exploration technique that has been applied to games. Previous work has shown that, when operating over a set of predefined Boolean pixel features, IW can achieve scores comparable to humans in almost real time when it comes to playing Atari 2600 games \cite{planning_with_pixels}. Our work focuses RRT over IW because RRT makes use of a continuous feature space, exactly the type learned by Pix2Mem.

In evolutionary computation, novelty search is the paradigm that eschews optimizing a given fitness function in favor of finding solutions that are different from those seen before. Techniques such as MAP-Elites \cite{mapelite} are explicitly designed to yield a large archive of solutions with Quality Diversity (QD) \cite{qualitydiversity}.

Rather than attempt to use intrinsically motivated reinforcement learning techniques directly as exploration strategies, our work focuses on using algorithms like RRT to reach new moments in the game and understand how they relate to one another. As in QD work, the emphasis is on the resulting archive.  For applications where training a policy is the desired goal, archives of interesting moments found by unrelated exploration strategies might provide a critical efficiency boost for learning~\cite{youtube}.

\section{Game Interface}

Rather than directly taking on the full complexity of a mobile game platform like Android, we consider a sequence of simpler game platforms. We start with Atari 2600 (a second-generation game console), the platform commonly used in recent deep reinforcement learning experiments. We end at Nintendo 64 (fifth-generation), a platform that modestly undershoots the technical specifications of low-end Android devices. The feature most distinguishing the platforms we consider from Android (aside from touch-screen controls) is the lack of networking capabilities. For better or for worse, many Android games are actually parts of larger distributed systems. This means that we can save and restore game states for earlier game contents (as required by MCTS-style forward models) if we are willing to pay the time and space cost of doing so. For Android, however, snapshotting the state of an active network connection would be unrealistic.

To interface with this broad range of game platforms, we selected the BizHawk\footnote{\url{http://tasvideos.org/Bizhawk.html}} emulator. Maintained by the tool-assisted speedrunning (TAS) community, BizHawk offers an extensive scripting interface which we can access via Python. Depending on the complexity of the emulated platform and the exploration strategy, one second of simulated gameplay may take more or less than one second of wall-clock time---exploration need not occur in real-time.

To define the space of possible actions for each platform, we recorded ourselves playing a small number of games for each. Most of our automated exploration strategies select actions (controller input states) only from those represented in this dataset.

\section{Exploration Strategies}

Each of our exploration strategies interacts with games in BizHawk over time. They may configure the controller state (defining which buttons are pressed) and advance the logic of the game. Strategies can decide whether any frame they experience is included in their output collection of moments (represented as screenshot and memory state snapshot pairs). Strategies may save and load any number of snapshots they like.

\subsubsection{Attract Mode}

The first and simplest exploration strategy we consider is associated with the concept of an attract mode. Attract modes are a feature of many games derived from arcade classics which played animations when idle to entice players to walk up and insert coins. Attract mode animations often show snippets of actual gameplay, profiles of characters, or cut-scenes revealing pieces of the main plot. In the case where gameplay is shown, the game is typically executing almost all the same code as in interactive play and a stored recording of representative play styles is seen. For knowledge extraction purposes, these demonstrations are as valid as those uncovered by interactive play.

Our attract mode strategy sets the no-buttons-pressed controller state and simply begins to advance game time by half a second each step. Although this strategy yields very interesting demonstrations of expert play and late-game content for some games, for many others it leaves the game sitting at an uninteresting screen that a human player would have quickly dismissed.

\subsubsection{Human Volunteers}

Our next strategy reliably yields samples of core gameplay behavior. We ask human volunteers to play the game while recording their input stream as a BizHawk movie. These BK2 files are a common medium for sharing interaction traces in the TAS community. In our experiments below, human data comes from players who have usually never played the specific game before. In some cases, our volunteers were not familiar with the language of the text on the screen. For certain games in our collection, there are numerous movies of expert play available for download from the TAS community.\footnote{\url{http://tasvideos.org/Movies.html}} For others, our team's first look at a game came from browsing the results of attract mode exploration.

To harvest a collection of moments from these pre-recorded movies, we play them back through BizHawk keeping frames spaced half a second apart as in the attract mode strategy. Our volunteers did not make use of BizHawk's save/load features to explore more thoroughly, and they were not given any specific instructions about what play style to use.

\subsubsection{Chaos Monkey}

Our volunteers noted that often random button mashing was good enough to stumble through menus they could not read. Our chaos monkey strategy is inspired by the ``UI/Application Exerciser Monkey'' distributed with the Android developer tools.\footnote{\url{https://developer.android.com/studio/test/monkey}} This tool injects a pseudo-random stream of input events (e.g. touch-screen actions) and system-level events (e.g.\ incoming phone calls) into an app under test in an effort to cause it to crash.

Our chaos monkey exploration strategy samples controller states according to a platform-wide distribution of controller states fit to our human gameplay collections. Actions are selected independent of the game, the contents of the display, and of previous actions. Similar to the attract mode strategy, we extract moments every half-second, holding the controller state between steps. 

Many of our games require the player to hit the controller's \textsc{start} button at least once during traversal of early game menus while not requiring this button later in play. As a result, our baseline chaos monkey strategy sometimes even triggers attract mode animations for failure to select this very rarely used input configuration in time. Certainly, there is much room for improvement.


\subsubsection{Rapidly-Exploring Random Trees}

Drawing on state-space exploration algorithms developed for robotics, this strategy uses the Rapidly-Exploring Random Trees algorithm \cite{rrt}. RRT is designed for exploring continuous state spaces of moderately-low dimensionality. In an application to robot arm movement, the state of the arm might be captured by joint angles. The ground-truth state of our videogames, however, spans a very high-dimensional discrete space:\footnote{Even though algorithms like IW can naturally operate on discrete representations, these algorithms would not make use of the knowledge implicitly represented in moment vectors to judge the significance of observed differences in display pixels or memory bytes.} the possible configuration of each byte of main memory and other subsystems. Even the space of display pixels is another high-dimensional discrete space. In response, we project screenshots into 256-dimensional moment vectors using the Pix2Mem deep convolutional network strategy \cite{moments}. Pix2Mem is a representation learning strategy in which the contents of memory are predicted from screenshot pixels. The bottleneck layer of this network serves as a compact, semantic representation of what is happening in the screenshot.

RRT works by growing a tree that captures how to reach points of the game's state space from some initial state (for us, the moment the  platform has booted with a game cartridge installed). Nodes in the tree are previously seen states, and edges (annotated with action sequences) describe how to get from one state to another. The algorithm continually picks a random goal location in the continuous moment vector space, finds which existing tree node is closest to it, and performs an action from that state in an attempt to get closer to the goal. The resulting action and state form a new tree edge.

This strategy makes extensive use of the ability to save and load game snapshots. To amortize the cost, the actions available to our RRT strategy operate over many frames. If snapshot loading were not available, states could be recreated by replaying actions along their path from the root node. In most applications of RRT, the selection of action to take next is parameterized by both the current state and the goal location. However, as in the chaos monkey strategy, our baseline RRT strategy selects a random human-demonstrated action and holds it for a half-second.

In our work RRT functions as both a source of data for knowledge extraction (producing screenshots and memory states on which to train) as well as an application for extracted knowledge (distances between moments in the game are judged by the similarity of their moment vectors). An experiment described later in this paper looks at the mutual benefit between Pix2Mem representation learning and RRT exploration.

\subsubsection{Meta Strategies}

The strategies introduced above offer many opportunities for local improvements. Without improving any individual strategy, we want to highlight how elementary strategies might be synergistically combined. When multiple strategies are applied to the same game, we ask what one can use of another's exploration results.

Rather than always starting from the same boot state, algorithmic strategies can be used to create branches off of human volunteer data. Similarly, RRT can be modestly adapted to produce rapidly-exploring random forests that branch off of state-space bookmarks placed by other exploration strategies. Rather than choosing actions independent of current game/state/goal, data from other strategies might be used to train a predictive model that selects a more relevant action by looking at the moment vectors of the current and goal states. One hybrid branching strategy is experimentally examined later in this paper.

Human volunteers, with access to a visualization of moments extracted by other strategies, might be able to find more interesting moments from which to start their own gameplay. This visualization might take the form of a tSNE \cite{tsne} plot of previously-seen moments (making use of the rich moment vector representation). Humans and algorithms together might be able to reach corners of the space that neither would have explored on their own in the same amount of total gameplay time. 

\subsection{Exploration Metrics}

How should we judge whether an exploration strategy is continuing to make progress or if one is making progress more efficiently than another? We would like to be able to directly observe how much territory is covered by a strategy by consulting a map. In our work, the space of rich, semantic moment vectors functions as this map.

By contrast to reinforcement learning research, we should emphasize that our goal is to assess the diversity of moments an exploration strategy has extracted, not to judge its gameplay behavior. Competent use of state saving/loading actions is highly relevant to exploration quality while not actually representing any in-game behavior at all.

\subsection{Measuring the Spread of Points}

Our preliminary exploration metrics are built on intuitions for how we might explore a low-dimensional, continuous space. Each moment extracted by an exploration strategy represents a point somewhere in this space. Intuitively, exploration strategies should produce a cloud of points that is well spread out in this space. How should we measure this?

One strategy is to draw an axis-aligned box (or hypercube) around all of the points and then to report the area (or hypervolume) of the box. As new points are explored and dropped into this space, the box can only grow when points fall outside the previous bounds, representing increasingly wider coverage. To account for degenerate boxes (which are well spread in some dimensions but not at all in others) we report the sum of the lengths of the sides of the box rather than their product. This is our \emph{bounding box sum} metric.

Another strategy pays less attention to extreme points and more to the distribution of points within the cloud. We propose to fit a (likely highly anisotropic) multi-dimensional Gaussian distribution to the points. The directions of maximal variation (derivable from the covariance matrix of the point data) play a similar role to the sides of the bounding box above, with variance being analogous to hypervolume. Again, to account for potentially degenerate distributions (for which the data is spread out only in a lower-dimensional subspace), we simply sum the eigenvalues of the covariance matrix (finding its nuclear norm) rather than forming their product (the determinant). This is our \emph{nuclear norm} metric.

By contrast with the bounding box sum metric, the nuclear norm metric may decrease as new points are explored if those points are relatively more concentrated than the initial set of points. This can happen for exploration strategies that spend a long time in one game mode or screen after initially traversing menus with rich visual variation.

\subsection{Embedding Game Moments into a Space}

Given that we can evaluate the spread of points in some space, how do we embed explored game moments into that space in the first place? A useful embedding function should place similar moments nearby one another and dissimilar moments far apart if our intuitions about the spread of points are to be effective.

Consider a game-specific embedding function for a game like \emph{Super Mario World} for the Super Nintendo Entertainment System. In one three-dimensional embedding, we might use one dimension to represent the game scene (an ordered index of game menus and then sequential level identifiers) and two more dimensions to represent the position of the player character within that scene. As an exploration strategy completes one scene and moves to the next, points on a new plane are discovered and discontinuous movement is represented in the positional dimensions. Although the truth of which scene we are in and where our character is in the scene could be computed from memory bytes, we consider a more general family of game-specific embeddings. In particular, we reuse the Pix2Mem embedding strategy (also used in the RRT strategy above). Zhan and Smith \shortcite{moments} showed that this representation was useful for identifying moments by scene as well as position within scene. This is our \emph{game-specific} embedding strategy.

To compare exploration across games and across game platforms, we cannot assume we have pre-established gameplay datasets available to train instances of Pix2Mem. Consider this game-independent three-dimensional embedding: the average RGB color of display pixels. When characters or other display elements move across colorful backgrounds by small amounts, the average color will only change a bit (if at all). Scene transitions, which might be marked by fades and flashes followed by new screens with different dominant colors, might be detected in this space. Three dimensions is clearly too small to be useful (particularly considering what flashes of all-white and all-black would do under the bounding box sum metric), but we can still recover the idea of a game-independent perceptual space. Recycling a specific\footnote{\url{https://keras.io/applications/#inceptionv3}} pre-trained neural network (the Inception-V3 network trained to convergence on ImageNet), we can embed screenshots into a 1000-dimensional space. Even though the output dimensions from this network are unrelated to the objects appearing in our videogame screenshots, it still seems to work in our experiments below. The all-white-all-black catastrophe for the bounding box sum metric, in particular, cannot occur with this embedding because the elements of the embedded vector are constrained to always add up to one. This is our \emph{game-independent} embedding strategy.

\begin{figure}
\centering
\includegraphics[width=\columnwidth]{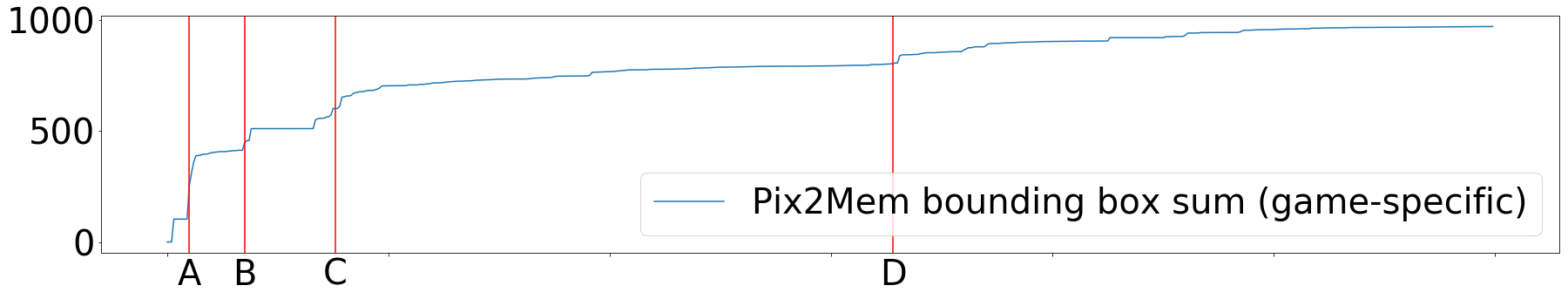}
\includegraphics[width=\columnwidth]{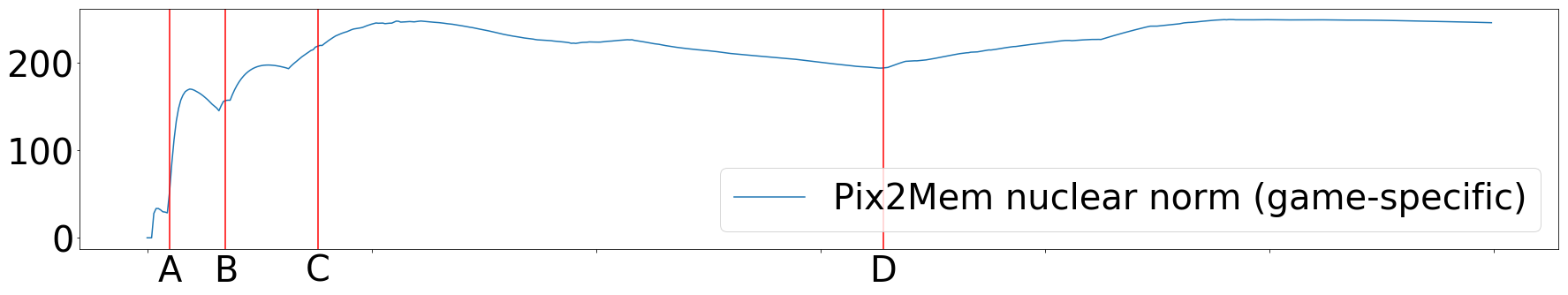}
\includegraphics[width=\columnwidth]{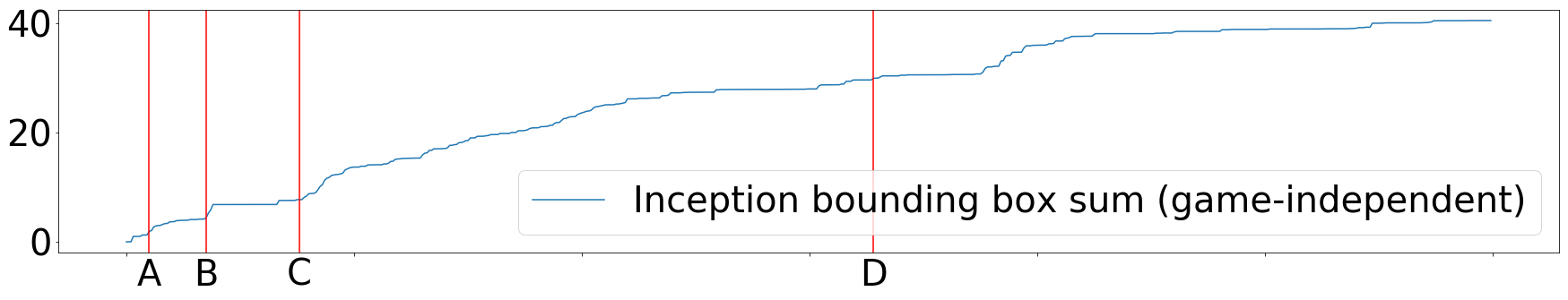}
\includegraphics[width=\columnwidth]{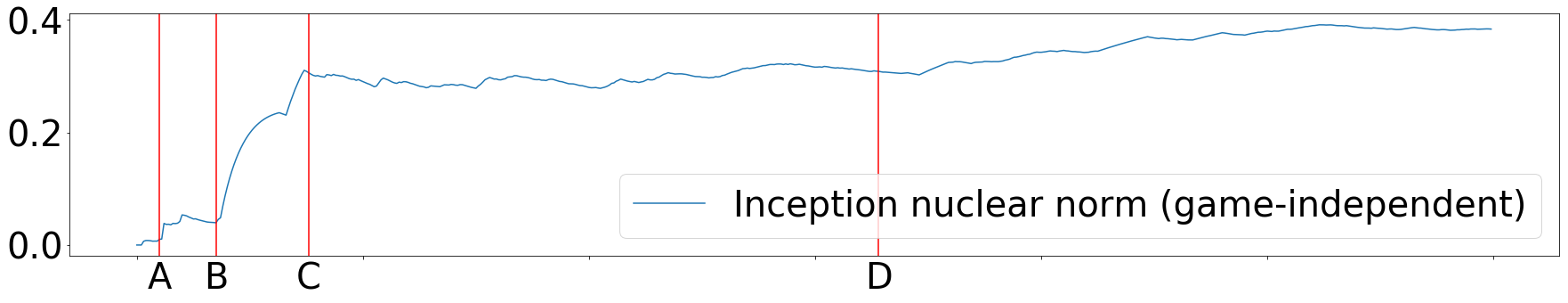}
\caption{Comparison of exploration quality metrics for a 5-minute run of human volunteer data in \emph{Super Mario World}. The marked moments represent approximately when (A) the main menu fades in, (B) the overworld map fades in, (C) first core (platformer) gameplay begins, and (D) first core gameplay completes.}
\label{fig:metrics}
\end{figure}

Fig.~\ref{fig:metrics} compares our two spread metrics with two embedding functions for a run of the human volunteer exploration strategy. Using the game-specific embedding (Pix2Mem trained on data from a longer human play session), each key moment (identified by visual inspection of the gameplay video) is associated with kinks in the graph for both spread metrics. For the game-independent (Inception) embedding, broad trends are somewhat preserved. The nuclear norm metric temporarily decreases for the game-specific embedding but not the game-independent embedding because the game-specific embedding knows our human player is still focused on one specific level for a long time while the game-independent metric is presumably more sensitive to background-art details that are scrolling by as the level progresses.

\section{Experiments}

This section describes experiments applying our exploration quality metrics to our elementary exploration strategies.

\subsection{Core Gameplay Coverage}

Can our automated exploration strategies make progress comparable to our human volunteers?

\begin{figure}
\centering
\includegraphics[width=\columnwidth]{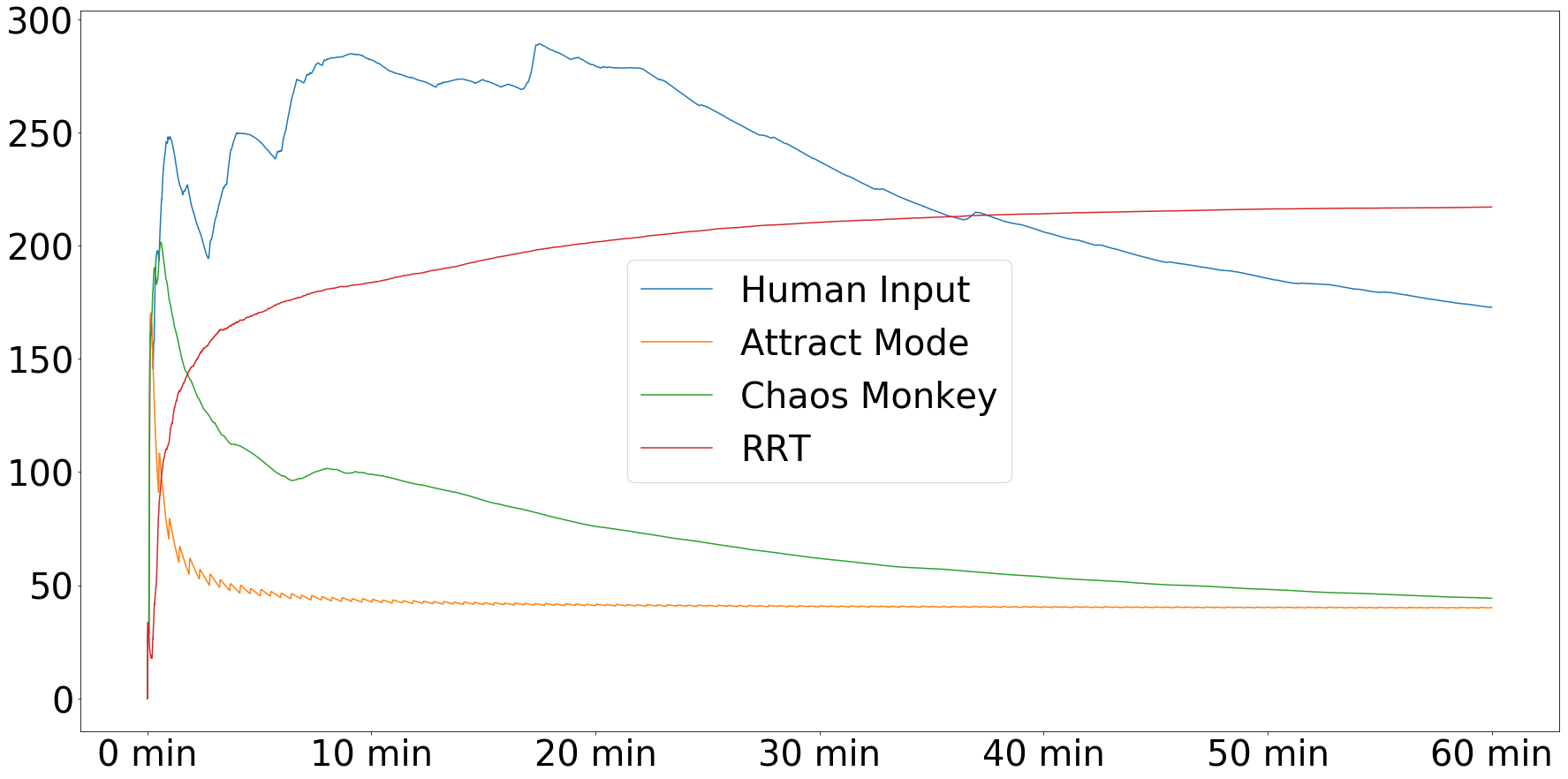}
\caption{Exploration progress (by Pix2Mem nuclear norm metric) on \emph{Super Mario World} by various strategies.}
\label{fig:basic}
\end{figure}

Fig.~\ref{fig:basic} visualizes the game-specific nuclear norm metric as a function of time for four exploration strategies. The attract mode strategy never starts core gameplay, however it does experience a short recording of example gameplay in the game's attract mode looping behind the main menu. Chaos monkey manages to start core gameplay almost as fast as the human volunteer, however it cannot sustain the diversity of human-demonstrated moments. RRT starts slow (often deciding to explore many moments selected from the animated transitions between scenes during which the player has no meaningful control) but eventually surpasses the spread from the other methods. Trends indicate that RRT would benefit from being able to explore for longer while the other strategies (including our human volunteer, one of the authors with imperfect gameplay skill) would not.

In a qualitative analysis of the screenshots extracted during exploration, we found that RRT could reach core gameplay in Super Mario World within two minutes when started from the game's boot state. When started from the first moment of core gameplay, RRT could finish the level within five minutes (less than 600 tree-expansion steps). Results like this encourage us to believe automatic exploration could be very useful for testing modern mobile games on a timescale that would be highly impractical for current deep reinforcement learning approaches. 

This experiment anchors our exploration metrics in a qualitative analysis of the modes and levels of one well-known game. In the remaining experiments, we do not perform a qualitative analysis as even the authors are not sufficiently familiar with the games in question to make a clear statement about what level of exploration would be enough for any given application.

\subsection{Bootstrapping Perception}

The experiment above executed RRT using a fixed embedding based on Pix2Mem trained on previous human gameplay for that specific game. Without falling back to a game-independent embedding (which would be reasonable anyway), can automatic exploration data itself be used as a source of training data for RRT's visual perception model?

In this experiment, we consider running RRT in the same game for 15 minutes at a time using different screenshot embedding functions. Our network starts untrained (weights randomly initialized). Although this embedding function has had no experience of the game under test (or any other game) it still supports automatic exploration, albeit at reduced efficiency. After the time is up, we incrementally train our network on the results of exploration run and discard the tree structure. We repeat this process of exploring and re-training several more times. Even though the final run of the algorithm still experiences only the same amount of total gameplay time and starts from the same initial state, it makes better progress through the game as a result of a better (game-specific) visual perception model.

\begin{figure}
\centering
\includegraphics[width=\columnwidth]{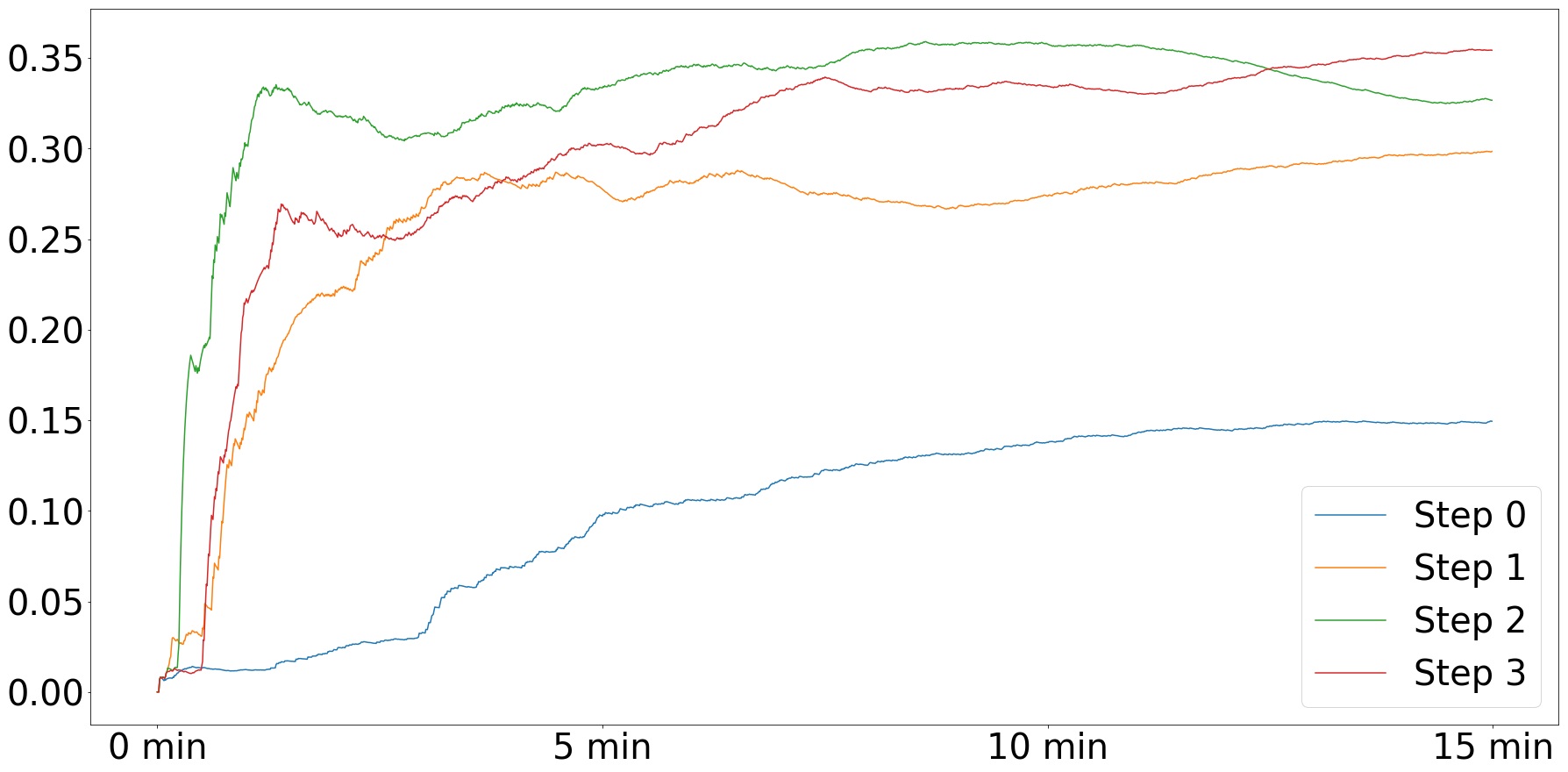}
\caption{RRT exploration progress (by Inception nuclear norm metric) for \emph{Super Mario World} using self-play to train the perceptual embedding network.}
\label{fig:bootstrap}
\end{figure}

Fig.~\ref{fig:bootstrap} shows the Inception\footnote{We used a game-independent metric for this experiment to avoid any confusion with metrics that might themselves be based on exploring this specific game} nuclear norm metric for four runs of RRT. Notice how more perceptual experience allows the fixed algorithm to explore both faster and deeper. The fact that the bulk of the gains are made after just the first 15 minutes of gameplay experience is promising for the use of game-specific embeddings even for games no human reviewers or testers have yet played (as in the app testing scenario).

\subsection{Cooperation between Strategies}

\begin{figure}
\centering
\includegraphics[width=\columnwidth]{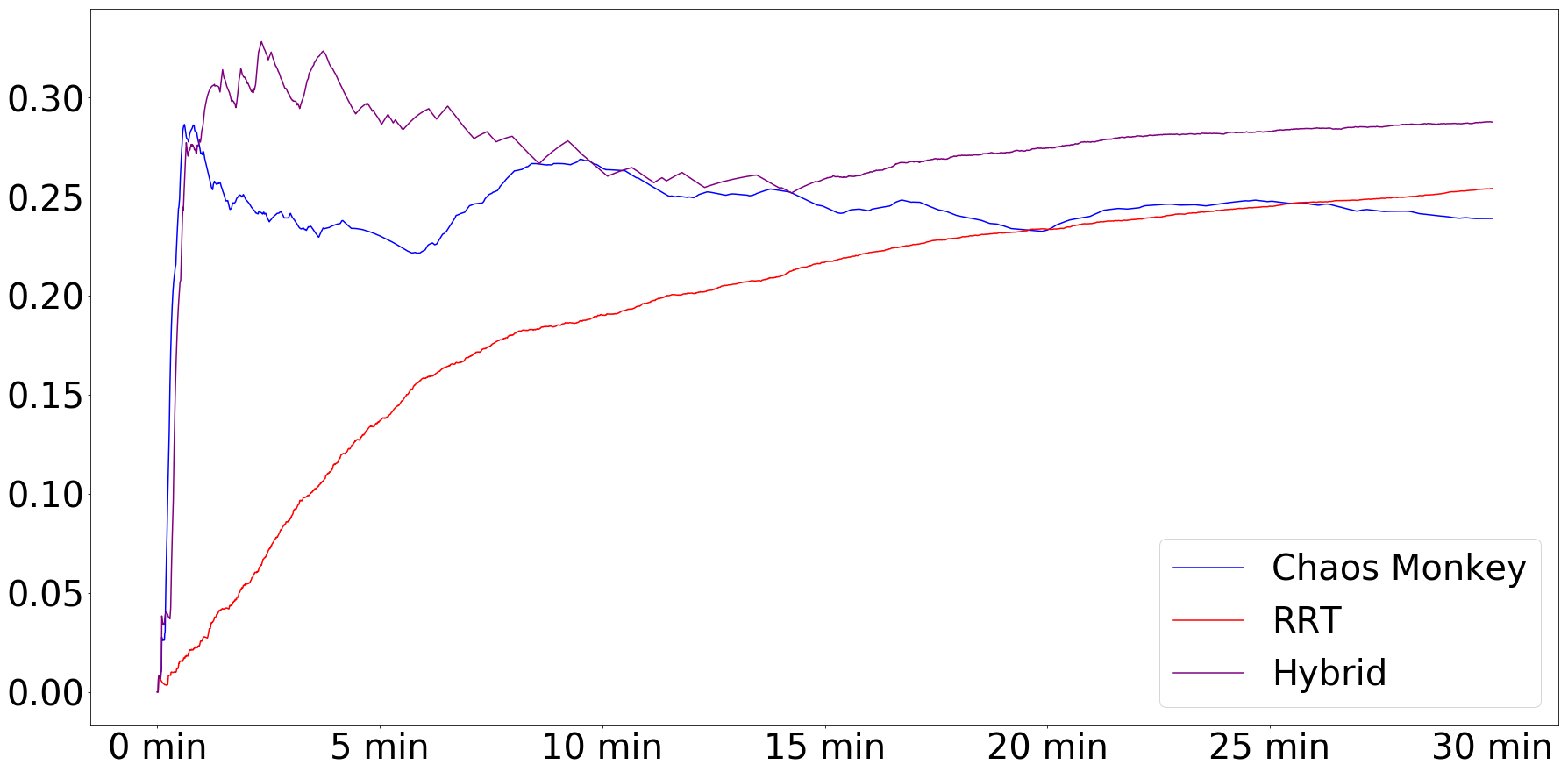}
\caption{Exploration progress (by Inception nuclear norm metric) for 30 total minutes of \emph{Super Mario World} gameplay using chaos monkey, RRT, and a combination where RRT explores from points first found by chaos monkey.}
\label{fig:cooperation}
\end{figure}

On a fixed budget, how should resources be allocated between different exploration strategies? Fig.~\ref{fig:cooperation} compares equal-time applications of chaos monkey and RRT with a hybrid strategy that switches from chaos monkey to RRT at the midpoint. The hybrid uses 100 randomly selected chaos monkey results to seed a forest of trees. As above, we show the Inception nuclear norm metric. Surprisingly, the hybrid strategy directly benefits from the quick start of chaos monkey while retaining the growth trend from RRT. The hybrid strategy covers more ground in equal time. In this case, only the set of seed states for exploration has been recycled, not a learned knowledge representation trained from chaos monkey's experience. In the previous experiment, we preserved the perceptual knowledge and not the concrete seed states.

\subsection{Performance across Games and Platforms}

Our final experiment compares the performance of our elementary exploration strategies across a number of commercial games selected from different BizHawk-supported game platforms. Because we use the same metric for each game (Inception nuclear norm after 10 minutes of simulated gameplay), scores may be compared across games and platforms. However, the range of scores depends both on game design issues (such as the degree to which the Inception network distinguishes visual variety in the game's display and how much variety can actually be experienced in just 10 minutes) and the utility of the exploration strategy. For randomized exploration strategies (chaos monkey and RRT), we average the score for 10 exploration runs.

\begin{figure}
\centering
\includegraphics[width=0.8\columnwidth]{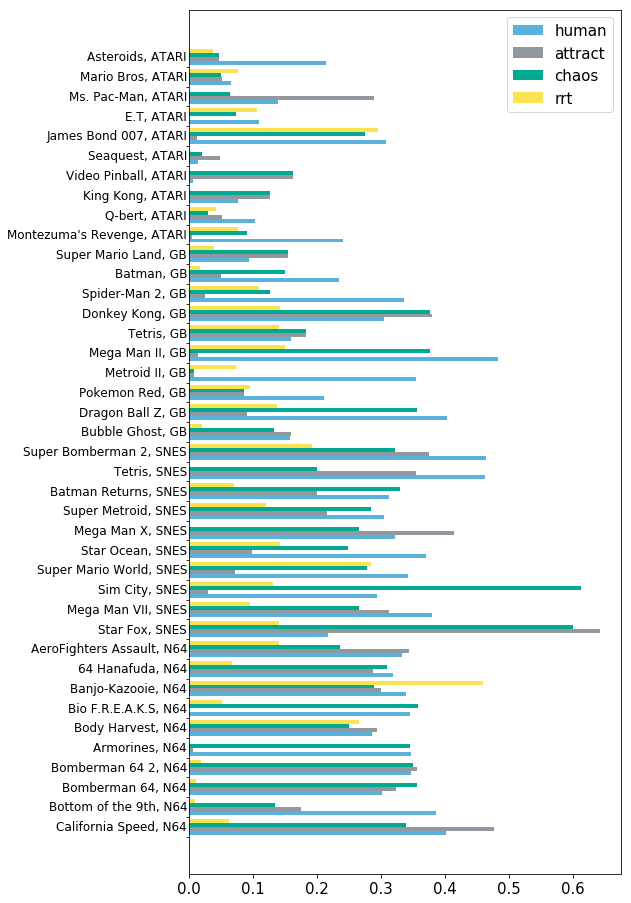}
\caption{Exploration progress after 10 minutes of simulated gameplay (by Inception nuclear norm metric) across games.}
\label{fig:across}
\end{figure}

Fig.~\ref{fig:across} samples games from Atari 2600 (128 bytes of memory), Game Boy (8 KiB), Super Nintendo (128 KiB), and Nintendo 64 (4 MiB). Notably, it is possible to surpass human exploration quality in some games, and no single strategy consistently out/under-performs others. From our bootstrapped perception experiment above, we know RRT has a slow start, particularly on the 10-minute scale and with a poor perceptual embedding (it uses the 1000-dimensional Inception embedding space in this experiment). We find these results encouraging for the possibility that modest improvements to our baseline strategies could yield human-comparable exploration on similar timescales. 

\section{Conclusion}

We introduce the problem of automatically exploring a game's state space with the goal of extracting a useful semantic map on timescales comparable to human playtesting efforts. We introduce elementary exploration strategies and demonstrate their application on a array of games and game platforms. To quantify exploration progress, we define four quality metrics that reveal different aspects of exploration patterns while exploiting pre-existing reference data when available. To compare exploration efficiency with human playtesters, we propose to compare quality metrics as a function of the total gameplay time used in exploration. Experimentally, we demonstrate that with a very modest amount of gameplay time, automatic exploration strategies can reach interesting gameplay moments (such as the completion of the first level in Super Mario World), bootstrap their own game-specific perception models, and leverage data produced by other exploration methods. Finally, we offer the first comparison of exploration progress across multiple games and platforms using a single platform-independent metric.


\begin{thebibliography}{}

\bibitem[\protect\citeauthoryear{Amalfitano \bgroup et al\mbox.\egroup
  }{2012}]{ripping}
Amalfitano, D.; Fasolino, A.~R.; Tramontana, P.; De~Carmine, S.; and Memon,
  A.~M.
\newblock 2012.
\newblock Using {GUI} ripping for automated testing of android applications.
\newblock In {\em Proceedings of the 27th IEEE/ACM International Conference on
  Automated Software Engineering},  258--261.
\newblock ACM.

\bibitem[\protect\citeauthoryear{Aytar \bgroup et al\mbox.\egroup
  }{2018}]{youtube}
Aytar, Y.; Pfaff, T.; Budden, D.; Paine, T.~L.; Wang, Z.; and de~Freitas, N.
\newblock 2018.
\newblock Playing hard exploration games by watching youtube.
\newblock {\em CoRR} abs/1805.11592.

\bibitem[\protect\citeauthoryear{Bandres, Bonet, and
  Geffner}{2018}]{planning_with_pixels}
Bandres, W.; Bonet, B.; and Geffner, H.
\newblock 2018.
\newblock Planning with pixels in (almost) real time.
\newblock {\em CoRR} abs/1801.03354.

\bibitem[\protect\citeauthoryear{Bauer and Popovic}{2012}]{treefrog}
Bauer, A.~W., and Popovic, Z.
\newblock 2012.
\newblock {RRT}-based game level analysis, visualization, and visual
  refinement.
\newblock In {\em Proceedings of the AAAI Conference on Artificial Intelligence
  in Interactive Entertainment}.

\bibitem[\protect\citeauthoryear{Bellemare \bgroup et al\mbox.\egroup
  }{2016}]{montezuma}
Bellemare, M.; Srinivasan, S.; Ostrovski, G.; Schaul, T.; Saxton, D.; and
  Munos, R.
\newblock 2016.
\newblock Unifying count-based exploration and intrinsic motivation.
\newblock In {\em Advances in Neural Information Processing Systems},
  1471--1479.

\bibitem[\protect\citeauthoryear{Browne \bgroup et al\mbox.\egroup
  }{2012}]{mcts}
Browne, C.~B.; Powley, E.; Whitehouse, D.; Lucas, S.~M.; Cowling, P.~I.;
  Rohlfshagen, P.; Tavener, S.; Perez, D.; Samothrakis, S.; and Colton, S.
\newblock 2012.
\newblock A survey of {M}onte {C}arlo tree search methods.
\newblock {\em IEEE Transactions on Computational Intelligence and AI in Games}
  4(1):1--43.

\bibitem[\protect\citeauthoryear{Burda \bgroup et al\mbox.\egroup
  }{2018}]{largescale}
Burda, Y.; Edwards, H.; Pathak, D.; Storkey, A.; Darrell, T.; and Efros, A.~A.
\newblock 2018.
\newblock Large-scale study of curiosity-driven learning.
\newblock In {\em arXiv:1808.04355}.

\bibitem[\protect\citeauthoryear{Chen \bgroup et al\mbox.\egroup
  }{2015}]{seconds}
Chen, K.; Wang, P.; Lee, Y.; Wang, X.; Zhang, N.; Huang, H.; Zou, W.; and Liu,
  P.
\newblock 2015.
\newblock Finding unknown malice in 10 seconds: Mass vetting for new threats at
  the {G}oogle-{P}lay scale.
\newblock In {\em 24th {USENIX} Security Symposium ({USENIX} Security 15)},
  659--674.
\newblock Washington, D.C.: {USENIX} Association.

\bibitem[\protect\citeauthoryear{Feng \bgroup et al\mbox.\egroup
  }{2014}]{apposcopy}
Feng, Y.; Anand, S.; Dillig, I.; and Aiken, A.
\newblock 2014.
\newblock Apposcopy: Semantics-based detection of {A}ndroid malware through
  static analysis.
\newblock In {\em Proceedings of the 22Nd ACM SIGSOFT International Symposium
  on Foundations of Software Engineering}, FSE 2014,  576--587.
\newblock New York, NY, USA: ACM.

\bibitem[\protect\citeauthoryear{Joulin \bgroup et al\mbox.\egroup
  }{2016}]{bagoftricks}
Joulin, A.; Grave, E.; Bojanowski, P.; and Mikolov, T.
\newblock 2016.
\newblock Bag of tricks for efficient text classification.
\newblock {\em CoRR} abs/1607.01759.

\bibitem[\protect\citeauthoryear{Kaltman \bgroup et al\mbox.\egroup
  }{2017}]{gisst}
Kaltman, E.; Osborn, J.; Wardrip-Fruin, N.; and Mateas, M.
\newblock 2017.
\newblock Getting the {GISST}: A toolkit for the creation, analysis and
  reference of game studies resources.
\newblock In {\em Proceedings of the 12th International Conference on the
  Foundations of Digital Games}, FDG '17,  16:1--16:10.
\newblock New York, NY, USA: ACM.

\bibitem[\protect\citeauthoryear{Kavraki \bgroup et al\mbox.\egroup
  }{1996}]{prm}
Kavraki, L.~E.; Svestka, P.; Latombe, J.-C.; and Overmars, M.~H.
\newblock 1996.
\newblock Probabilistic roadmaps for path planning in high-dimensional
  configuration spaces.
\newblock {\em IEEE transactions on Robotics and Automation} 12(4):566--580.

\bibitem[\protect\citeauthoryear{LaValle}{1998}]{rrt}
LaValle, S.~M.
\newblock 1998.
\newblock Rapidly-exploring random trees: A new tool for path planning.
\newblock Technical Report TR 98-11, Computer Science Department, Iowa State
  University.

\bibitem[\protect\citeauthoryear{Maaten and Hinton}{2008}]{tsne}
Maaten, L. v.~d., and Hinton, G.
\newblock 2008.
\newblock Visualizing data using t-sne.
\newblock {\em Journal of machine learning research} 9(Nov):2579--2605.

\bibitem[\protect\citeauthoryear{Merrick and Maher}{2009}]{curious}
Merrick, K.~E., and Maher, M.~L.
\newblock 2009.
\newblock {\em Motivated reinforcement learning: curious characters for
  multiuser games}.
\newblock Springer Science \& Business Media.

\bibitem[\protect\citeauthoryear{Mikolov, Yih, and Zweig}{2013}]{wordanalogy}
Mikolov, T.; Yih, W.-t.; and Zweig, G.
\newblock 2013.
\newblock Linguistic regularities in continuous space word representations.
\newblock In {\em Proceedings of the 2013 Conference of the North American
  Chapter of the Association for Computational Linguistics: Human Language
  Technologies},  746--751.

\bibitem[\protect\citeauthoryear{Mnih \bgroup et al\mbox.\egroup }{2015}]{dqn}
Mnih, V.; Kavukcuoglu, K.; Silver, D.; Rusu, A.~A.; Veness, J.; Bellemare,
  M.~G.; Graves, A.; Riedmiller, M.; Fidjeland, A.~K.; Ostrovski, G.; et~al.
\newblock 2015.
\newblock Human-level control through deep reinforcement learning.
\newblock {\em Nature} 518(7540):529.

\bibitem[\protect\citeauthoryear{Mouret and Clune}{2015}]{mapelite}
Mouret, J.-B., and Clune, J.
\newblock 2015.
\newblock Illuminating search spaces by mapping elites.
\newblock {\em arXiv preprint arXiv:1504.04909}.

\bibitem[\protect\citeauthoryear{Osborn, Summerville, and Mateas}{2017a}]{agdl}
Osborn, J.~C.; Summerville, A.; and Mateas, M.
\newblock 2017a.
\newblock Automated game design learning.
\newblock In {\em 2017 IEEE Conference on Computational Intelligence and Games
  (CIG)},  240--247.

\bibitem[\protect\citeauthoryear{Osborn, Summerville, and
  Mateas}{2017b}]{mappy}
Osborn, J.; Summerville, A.; and Mateas, M.
\newblock 2017b.
\newblock Automatic mapping of {NES} games with mappy.
\newblock In {\em FDG '17 Proceedings of the 12th International Conference on
  the Foundations of Digital Games}.
\newblock ACM.

\bibitem[\protect\citeauthoryear{Pugh, Soros, and
  Stanley}{2016}]{qualitydiversity}
Pugh, J.~K.; Soros, L.~B.; and Stanley, K.~O.
\newblock 2016.
\newblock Quality diversity: A new frontier for evolutionary computation.
\newblock {\em Frontiers in Robotics and AI} 3:40.

\bibitem[\protect\citeauthoryear{Shen, Chien, and Hung}{2014}]{malware}
Shen, Y.-C.; Chien, R.; and Hung, S.-H.
\newblock 2014.
\newblock Toward efficient dynamic analysis and testing for {A}ndroid malware.
\newblock {\em IT CoNvergence PRActice (INPRA)} 2(3):14--23.

\bibitem[\protect\citeauthoryear{Zhan and Smith}{2018}]{moments}
Zhan, Z., and Smith, A.~M.
\newblock 2018.
\newblock Retrieving game states with moment vectors.
\newblock In {\em Proceedings of the AAAI 2018 Workshop on Knowledge Extraction
  from Games}.

\bibitem[\protect\citeauthoryear{Zhang \bgroup et al\mbox.\egroup
  }{2018}]{crawling}
Zhang, X.; Zhan, Z.; Holtz, M.; and Smith, A.~M.
\newblock 2018.
\newblock Crawling, indexing, and retrieving moments in videogames.
\newblock In {\em FDG '18 Proceedings of the International Conference on the
  Foundations of Digital Games}.
\newblock ACM.

\bibitem[\protect\citeauthoryear{Zook, Harrison, and Riedl}{2015}]{zook}
Zook, A.; Harrison, B.; and Riedl, M.
\newblock 2015.
\newblock Monte-carlo tree search for simulation-based play strategy analysis.
\newblock In {\em Proceedings of the 10th International Conference on the
  Foundations of Digital Games, {FDG} 2015, Pacific Grove, CA, USA, June 22-25,
  2015}.

\end{thebibliography}

\bibliographystyle{aaai}
\end{document}